\documentclass[twocolumn,showpacs,floatfix,preprintnumbers,amsmath,amssymb,prl]{revtex4}
\usepackage{epsf}
\usepackage{graphicx}
\usepackage{dcolumn}
\usepackage{bm}

\begin{document}

\preprint{To be presented at 14$^{\rm th}$ Int. Symposium ``Nanostructures: Physics and Technology'',
                St.~Petersburg, Russia, June 26--30, 2006}

\title{Optics of Charged Excitons in Quantum Wells: \\
Free versus Donor-bound Complexes}

\author{A.~B.~Dzyubenko$^{1,2}$}
\author{D.~A.~Cosma$^{1}$}
\author{T.~D.~Kelly~II$^{1}$}
\author{A.~R.~Todd$^{1}$}
\author{A.~Yu.~Sivachenko$^{3}$}

\affiliation{
$^{1}$Department of Physics, California State University at Bakersfield, Bakersfield, CA 93311, USA \\
$^{2}$General Physics Institute, Russian Academy of Sciences,
            Vavilova 38, Moscow 119991, Russia  \\
$^{3}$ Ariadne Genomics Inc., 9700 Great Seneca Highway,
             Rockville, MD 20850, USA
}

\email{adzyubenko@csub.edu}

\date{\today}

\begin{abstract}
We theoretically study localization of quasi-two-dimensional negatively
charged excitons $X^-$ on isolated charged donors in magnetic fields.
We consider donors located in a barrier at various distances $L$
from the heteroboundary as well as donors in the quantum well.
We establish how many different singlet $X_s^-$
and triplet $X_t^-$  bound states a donor ion $D^+$  can support in
magnetic fields  $B>6$~T. We find several new bound states,
some of which have surprisingly large oscillator strengths.
\end{abstract}



\maketitle

\section{Introduction}

Optical signatures of spin-singlet  $X_s^-$
and spin-triplet  $X_t^-$ charged excitons are commonly
observed in semiconductor nanostructures in magnetic fields.
Despite the status of $X^-$ as one of the simplest few-body systems
with Coulomb interactions and a large amount of experimental
and theoretical work, some important issues remain unresolved.
One of these issues is the degree of localization of charged excitons
in realistic quantum wells (QW's) and how localization of $X^-$
manifests itself in optics \cite{Volkov1998,Dzyub2000}.

In this work, we discuss exact selection rules that
govern transitions of charged excitons in magnetic fields.
We also demonstrate, on a quantitative level, how these
selection rules work when applied to free $X^-$
and donor-bound $(D^+,X^-)$ charged excitons; the latter
can also be considered as excitons bound
to a neutral donor $(D^0,X)$.

\section {Classification of states and  selection rules}

Classification of states of free charged electron-hole
complexes in magnetic fields is based on magnetic translations
and the axial symmetry about the magnetic field axis \cite{Dzyub2000}.
The corresponding orbital quantum numbers are the oscillator
quantum number $k = 0, 1, 2, \ldots$ and
the total angular momentum projection on the $z$-axis, $M$.
The former has the meaning of the mean squared distance
to the orbit guiding center; there is an infinite-fold Landau
degeneracy in $k$.
Each family of degenerate $X^-$ states starts with its parent $k=0$
state that has some specific value of $M$ that follows from
the solution of the Schr\"{o}dinger equation. Degenerate daughter
states $k =  1, 2,  \ldots$ have values $M-1, M-2, \ldots$
for the total angular momentum projection.
Selection rules for interband transitions are
$\Delta M = 0$  and $\Delta k = 0$ and lead to the following
results:  Photoluminescence (PL) of a free $X^-$
must leave an electron in a LL with the number $n$ equal to the angular
momentum $M$ of the parent state,
$X^- \rightarrow \hbar\omega_{X^-} + e^{-}_{n=M}$.
Therefore, (i) families of $X^-$ states that start with $M < 0$
are dark in PL and (ii) shake-ups to multiple Landau levels (LL's)
are strictly prohibited \cite{Dzyub2000}.

The presence of a donor ion $D^+$ breaks the translational symmetry,
lifts the degeneracy in $k$, and makes many of the previously prohibited
transitions allowed. Let us discuss spectroscopic consequences
of the remaining axial symmetry for a donor-bound state $(D^+,X^-)$
with angular momentum $M$ and wavefunction
$\Psi_{M} ({\bf r}_{e1},{\bf r}_{e2};{\bf r}_h)$.
The dipole matrix element for interband transition to a final
electron state $\phi_{m_f}({\bf r})$ with angular momentum projection $m_f$ is
\begin{equation}
    \label{dipole}
       f \sim \left| p_{\rm cv} \int \!\! d{\bf r} \int \!\! d{\bf r'} \,
                   \phi_{m_f}^*({\bf r}) \,
                   \Psi_M({\bf r},{\bf r'};{\bf r'}) \right|^2 \sim \delta_{M,m_f}  \: .
\end{equation}
Conservation of angular momentum $M = m_f$  can be satisfied
for a number of final states $\phi_{m_f}({\bf r})$ belonging to different LL's.
Therefore, shake-up processes become allowed in PL. More than that,
PL of $(D^+,X^-)$ states with $M > 0$ must proceed
via shake-ups to higher LL's  \cite{Dzyub1993}. 
This is because
electron states with angular momenta $m_f  = M > 0$ are only
available in $n=M$ or higher LL's. Note that the shake-up
processes are due to the Coulomb induced admixture of
LL's and are suppressed in strong fields as $B^{-2}$.

\section {Numerical approach}

We obtain the energies and wavefunctions
of the $(D^+,X^-)$ complexes and of free $X^-$ excitons
by diagonalization of the interaction Hamiltonian using a
complete basis of states compatible with both axial and
electron permutational symmetries. The basis states are
constructed out of the in-plane wavefunctions in LL's
and size quantization levels in a QW with proper
symmetrization for triplet and singlet states.
We consider up to $2\times 10^5$ basis states and,
by applying an adaptive scheme, we choose out of these
about $6 \times 10^3$ states to be diagonalized in the Hamiltonian matrix.
A stability for the donor-bound charged exciton is determined
with respect to its dissociation to a neutral donor
and a free exciton, $(D^+,X^-) \rightarrow D^0 + X$.
Accordingly, a binding energy of a stable
$(D^+,X^-)$ complex is defined as the energy difference
between the total Coulomb energies
\begin{equation}
  \label{Dxb}
        E^{\rm b}_{(D^+,X^-)} =  E^{\rm Coul}_{D^0} + E^{\rm Coul}_{X}
                                                  - E^{\rm Coul}_{(D^+,X^-)}  > 0 \: ,
\end{equation}
with $D^0$ and $X$ being in their ground states.
Binding energy (\ref{Dxb}) determines the energy difference
between the positions of a neutral exciton and a donor-bound charged exciton PL emission
lines,
$\hbar\omega_{X} -  \hbar\omega_{(D^+,X^-)}= E^{\rm b}_{(D^+,X^-)}$.

\section {Results and discussion}

The calculated binding energies of the various
charged exciton states in a 100~{\AA} GaAs QW as functions of the distance $L$
to the donor ion $D^+$ are shown in Fig.~1.
We estimate the accuracy in binding energies to be of the order of $\pm 0.1$~meV.

The limiting case $L = \infty$ corresponds to free charged excitons $X^-$.
There are three documented bound states in this limit:
the bright singlet $X^-_s$ with $M=0$, the dark triplet $X^-_{td}$ with
$M=-1$, and the bright triplet $X^-_{tb}$ with $M=0$
(see \onlinecite{Dzyub2000,Wojs2000,Riva2001} and references therein).
We found two new bound states: the second dark
triplet state with $M=-1$, labeled $X^-_{td2}$ in Fig.~1,
and the second bright singlet state with $M=0$, labeled
$X^-_{s2}$. These states are very weakly bound
and will be discussed in more detail elsewhere.

Our results show that the parent bright singlet
state $X^-_s$ with $M=0$ remains always bound.
Its binding energy initially decreases with decreasing $L$,
reaches its minimum when the donor $D^+$
is very close to the heteroboundary, and then increases again.
We interpret this as an indication toward a rearrangement
of the type of binding in the singlet $(D^+,X_s^-)$ state:
At very large distances $L$,
the donor ion binds $X_s^-$ as a whole, barely affecting its internal
structure. In the opposite limit of an in-well donor, the interaction of
electrons with the $D^+$ is stronger than that with the hole.
The donor-bound complex formed in this case is better described as
an exciton bound to a neutral donor $(D^0,X)$.

Notice a systematic change in the dipole transition matrix elements
$f$ in Fig.~1: as the binding energy of a complex decreases,
its spatial extent increases leading to the increase in $f$.
This is consistent with the notion of ``giant oscillator strengths''  \cite{Rashba1962}.
%
%
\begin{figure}
\leavevmode \centering{\epsfbox{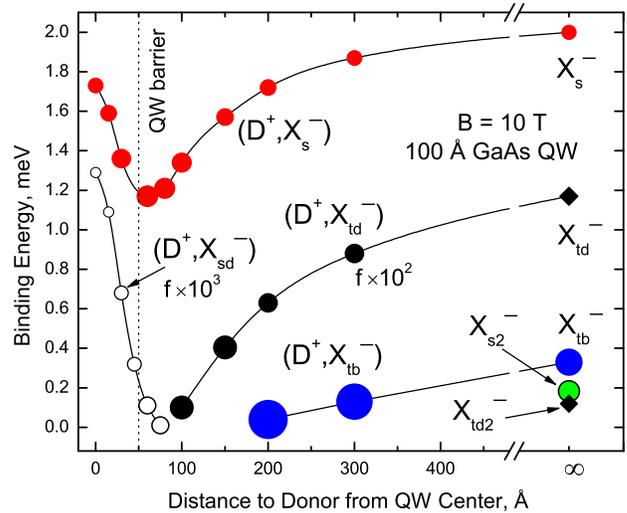}}
\caption[h]{Binding energies of charged excitons $X^-$
in a 100~\AA\ GaAs/Al$_{0.3}$Ga$_{0.7}$As QW at $B=10$~T.
Sizes of the dots are proportional to the interband dipole
transition matrix elements $f$.
The solid diamonds designate dark $X^-$ states.}
\end{figure}
%
%

We found just one state that only exists in the presence of the $D^+$
and does not have its free $L = \infty$ counterpart:
the dark singlet state $(D^+,X^-_{sd})$ with $M = 1$.
It only becomes bound when the $D^+$ is located in a QW or very near to it.
This is also the only donor-bound state that remains bound
in the strictly 2D high-field limit in symmetric electron-hole systems \cite{Dzyub1993}.
According to  (\ref{dipole}), the PL from this state goes mostly
via shake-ups to $n=1$ electron LL. As a result,
the dipole transition matrix elements $f$ shown in Fig.~1 are very small.
%
%
\begin{figure}[h]
\leavevmode \centering{\epsfbox{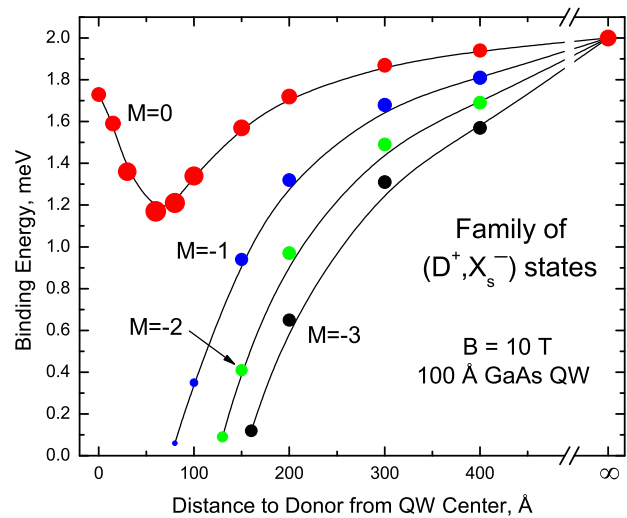}}
\caption{Lifting of the Landau degeneracy in the
family of bright singlet states $X_s^-$.
The states are characterized by different total angular momentum projections
$M=0, -1, -2, \ldots $ and all are optically active.
 }
\end{figure}
%
%

In contrast to singlet states, the dark  $X^-_{td}$ and bright $X^-_{tb}$
triplet states survive only for sufficiently large distances $L$
to the donor ion $D^+$ (Fig.~1). This is because electrons
in triplet states cannot simultaneously occupy the $s$-state in the lowest LL
and, therefore, it is difficult to find a configuration with
optimized electron-donor interactions.
Notice the finite oscillator strengths for the PL from the donor-bound
complex $(D^+,X_{td}^-)$ originating from the dark triplet state.

We stress that each free $X^-$ state gives rise to a family
of degenerate states; only the evolution of the parent $X^-$  states is shown in Fig.~1.
The degeneracy in the in-plane position of the guiding center (quantum number $k$)
is lifted in the presence of the donor ion $D^+$.
Figure~2 demonstrates this for the family of singlet bright states $X^-_s$.
When the distance to the donor $L$ decreases, all but one state (with $M=0$)
become one by one unbound. This leads to a number of optically active states
with large oscillator strengths.

In conclusion, we have shown there is a multitude of donor-bound
$X^-$ states that may exhibit relatively weak dependencies of
binding energies and oscillator strengths on positions of remote donors.
Our results may be relevant for explanation
of the PL from the dark triplet state $X_{td}^-$,
of the multiple PL peaks observed in different experiments,
and of the $X^-$ shake-ups in PL.

\begin{acknowledgments}
This work is supported in part by NSF grants
DMR-0203560 and DMR-0224225, and by a College Award of Cottrell
Research Corporation.
\end{acknowledgments}

 

\begin{thebibliography}{8}
\itemsep-2pt

\mbox{}
\vspace{0.1cm}
\mbox{}
	
\bibitem{Volkov1998}
O.~V.~Volkov, V.~E.~Zhitomirskii,
I.~V.~Kukushkin, K.~von Klitzing, and K.~Eberl,
{\em Pis'ma ZhETF\/} {\bf 67}, 707 (1998)
[{\em JETP Lett.} {\bf 67}, 744  (1998)].

\bibitem{Dzyub2000}
A.~B.~Dzyubenko and A.~Yu.~Sivachenko,
{\em Phys. Rev. Lett.} {\bf 84}, 4429 (2000).

\bibitem{Dzyub1993}
A.~B.~Dzyubenko,
{\em Phys. Lett. A} {\bf 173}, 311 (1993).

\bibitem{Wojs2000}
A.~W\'ojs, J.~J.~Quinn, and P.~Hawrylak,
{\em Phys. Rev. B} {\bf 62}, 4630 (2000).

\bibitem{Riva2001}
C.~Riva, F.~M.~Peeters, and K.~Varga,
{\em Phys. Rev. B} {\bf 63}, 115302 (2001).

\bibitem{Rashba1962}
E.~I.~Rashba and G.~E.~Gurgenishvili,
{\em Fiz.~Tverd.~Tela} {\bf 4}, 1029 (1962)
[{\em Sov.~Phys.~Solid State} {\bf 4}, 759 (1962)].


\end{thebibliography}
\end{document}